\documentclass[11pt,twocolumn]{article}
\usepackage[T1]{fontenc}
\usepackage{ae,aecompl}

\usepackage{graphicx}	
\usepackage{amssymb}	
\usepackage{longtable}

\usepackage[utf8]{inputenc}
\usepackage{pdflscape}
\usepackage{lscape}
\usepackage{booktabs}
\usepackage{csquotes}
\usepackage{color, colortbl}
\usepackage{amstext}
\usepackage{wasysym}
\usepackage{multirow}
\usepackage{natbib}
\usepackage{ulem}

\def\degr{\hbox{$^\circ$}}

\def\fdg{\hbox{$.\!\!^\circ$}}

\usepackage{t1enc}
\usepackage{times}

\usepackage{graphicx}
\usepackage{footnote}
\usepackage{pifont}
\usepackage{lscape}
\usepackage{booktabs}

\usepackage{threeparttablex}
\definecolor{LightGray}{gray}{0.9}
\definecolor{LightGray1}{gray}{0.8}
\definecolor{pad}{rgb}{0.06,0.15,0.76}
\definecolor{PAD}{rgb}{0.06,0.15,0.76}

\usepackage{authblk}
\usepackage{blindtext}
\usepackage[margin=1.6cm]{geometry}

\usepackage{hyperref}

\title{Kruger~60 as a home system for 2I/Borisov -- a case study}

\author[1]{Piotr A. Dybczyński\thanks{Corresponding Author: dybol@amu.edu.pl}}
\author[2]{Małgorzata Królikowska \thanks{mkr@cbk.waw.pl}}
\author[1]{Rita Wysoczańska \thanks{rita.wysoczanska@amu.edu.pl}}
\affil[1]{Astronomical Observatory Institute, Faculty of Physics, A.~Mickiewicz University, S{\l}oneczna 36, 60-286 Pozna\'{n}, Poland}
\affil[2]{Space Research Centre of the Polish Academy of Sciences, Bartycka 18A, 00-716 Warsaw, Poland}

\begin{document}
\label{firstpage}
\maketitle

\begin{abstract}
For the second discovered interstellar comet 2I/Borisov we searched for a candidate for its home system. We will never be sure which star or stellar system does this comet come from but obtaining a very small relative velocity and a promisingly small miss-distance, when tracing the motion of 2I/Borisov back in time in its movement through the space, makes an encountered body a good candidate for a source of this comet. In our long-standing project on studying Oort spike comets dynamics, we recently updated a list of potential stellar perturbers of cometary motion. This list was checked against a past, close and slow encounter with 2I/Borisov. Only one object from among 647~stars or stellar systems in our list, a double star Kruger~60, appeared as a potential candidate for the origin of this comet. However, a detailed analysis of this system's radial velocity uncertainty influence on our result showed, that the probability that Kruger~60 is a home system of 2I/Borisov is  small. Finally, the usage of a new, unpublished systemic radial velocity of Kruger~60 system practically ruled out this possibility.
\end{abstract}

\section{Introduction}

An interstellar comet 2I/Borisov, previously known as C/2019~Q4, was discovered on August~30 by an amateur astronomer Gennady Borisov at MARGOT observatory (Crimea). After this discovery was posted on the Minor Planet Center (MPC) Potential Comet Confirmation page\footnote{ https://www.minorplanetcenter.net/iau/NEO/pccp{\_}tabular.html}, several confirmations from different observatories appeared. Finally, the discovery and a nature of the object (cometary and interstellar) was officially announced in MPC electronic circular MPEC 2019-R106\footnote{https://minorplanetcenter.net/mpec/K19/K19RA6.html} \citep{2019MPEC-R106}. At that moment 145 positional observations were collected and the best fit to data gave a hyperbolic orbital solution with unusually large eccentricity of 3.08. In contrast to the previously discovered interstellar body, 1I/'Oumuamua, a cometary activity is evident from the beginning. At the moment of its discovery 2I/Borisov  was 2.99\,au from the Sun and 3.72\,au from the Earth on its inbound leg of the trajectory. The comet perihelion passage date is 2019, December~8 at a distance of 2.01\,au from the Sun.

One of the obvious questions after such a discovery is where does this object come from. When searching for a home system of an interstellar body one should look for a past close proximity with a star or stellar system that occurred at a very small relative velocity. Recent great progress in obtaining positional and kinematic data on nearby stars resulting (among others) from the Gaia mission \citep{Gaia_mission:2016} allowed us to update our list of potential stellar perturbers of cometary motion, see \cite{RPM:2019} (this list will be made publicly available in the near future). Such a list is a good starting point for a home system search, however other, more distant stars, can be also checked, but providing less accurate results since they are more distant in space and time, see for example \cite{Hallatt:2019}. 

Basing on the data available at that moment we announced, in the first version of this report, \citep[][v1]{dyb-kro-wys-v1:2019} that a double star Kruger~60 seems to be a good candidate for a home system of this comet and the only one from our list of stars or stellar systems. Now we refine our conclusions using much more positional data of 2I/Borisov (including prediscovery observations), adding Kruger~60 radial velocity uncertainty impact analysis and using the new, unpublished value of it.

In Section~\ref{sect:comet_orbit} we describe an updated orbit determination for this comet, including an attempt to determine non-gravitational effects. Sections~\ref{sect:stars} and \ref{sect:Kruger} present results of our search for the 2I/Borisov home system. In Section~\ref{sect:past_proximity} we analyze in detail the past approach of this comet with the double system Kruger~60, including the influence of poorly known radial velocity of this system on our result. Section~\ref{sect:conclusions} brings  conclusions.

\begin{table*}
	\caption{Barycentric, original orbits of 2I/Borisov obtained at 250\,au before entering the planetary zone. Orbits (n5), (n6) and (JPL2) represent NG~solutions while (a6), (b5) and (c5) are purely gravitational, as well as both orbits based on the MPC results. Equator and ecliptic of J2000 is used, osculation epoch of all orbits is 1983~Sept.\,23. For JPL original solutions we copied uncertainties for osculating orbit presented in JPL Small Body Database Browser assuming that the uncertainties of original orbital elements would be similar. The uncertainties of the MPC solution are not provided at MPC WebPage. }
	\label{table:comet_orbital_elements_original}
	\centering
	{{
			\setlength{\tabcolsep}{5.1pt} 
			\begin{tabular}{crrrrrrr}
				\hline 
&&&&&&& \\
solution  &    $T_{\rm ori}$ &    $q_{\rm ori}$~[au]    &    $e_{\rm ori}$        &  $\omega _{\rm ori}$ [\degr]  &  $\Omega _{\rm ori}$ [\degr]   &   $i_{\rm ori}$ [\degr]  & $1/a_{\rm ori}$~[au$^{-1}$] \\
&&&&&&& \\
\hline
&&&&&&& \\
\multicolumn{8}{c}{Data arc:  2019 Aug. 30 -- 2019 Sep. 23, 548 obs.} \\
&&&&&&& \\
a6 &  2019 Dec.\,8.74714 &     2.014522 &     3.364120 &   209.08984 &   308.11074 &  44.03931 & $-$1.173539 \\
   &         $\pm$.05738 & $\pm$.002251 & $\pm$.010252 & $\pm$.05002 & $\pm$.02070 &$\pm$.02082 &$\pm$.003779 \\
&&&&&&& \\
\multicolumn{8}{c}{Data arc:  2019 Aug. 30 -- 2019 Oct. 21, 1459 obs.; data taken from MPC and JPL WebPages on Oct.~30} \\
&&&&&&& \\ 
b5 &  2019 Dec.\,8.73613 &     2.014802 &     3.366600 &   209.08605 &   308.11435 &   44.03628 & $-$1.174607 \\
   &         $\pm$.00363 & $\pm$.000138 & $\pm$.000729 & $\pm$.00279 & $\pm$.00135 &$\pm$.00127 &$\pm$.000281 \\
&&&&&&& \\
MPC1 &  2019 Dec.\,8.76852 &     2.013615 &     3.360724 &   209.11148 &   308.10296 &   44.04725 & $-$1.172381 \\
&&&&&&& \\
JPL1 &  2019 Dec.\,8.77861 &     2.013275 &     3.359597 &   209.12096 &   308.10061 &   44.04974 & $-$1.172019 \\
     &        $\pm$.00960 & $\pm$.000368 & $\pm$.001937 & $\pm$.00745 & $\pm$.00361 &$\pm$.00341 &             \\
&&&&&&& \\
\multicolumn{8}{c}{Data arc:  2018 Dec. 13 -- 2019 Nov.  2, 1711 obs.; data taken from MPC and JPL WebPages on Nov.~5} \\
&&&&&&& \\
c5  &  2019 Dec.\,8.755747&     2.01406219 &    3.36321624 &  209.102714 &  308.107625 &   44.043060 &  $-$1.173358 \\
      &       $\pm$.000506& $\pm$.00001693 &$\pm$.00008010 &$\pm$.000405 &$\pm$.000150 &$\pm$.000152 & $\pm$.000030 \\
&&&&&&& \\
MPC2 & 2019 Dec.\,8.757111&     2.01405224 &    3.36073172 &  209.103294 &  308.107419 &   44.043159 & $-$1.172130 \\
&&&&&&& \\
n5  &  2019 Dec.\,8.754710&     2.01402871 &    3.36325634 &  209.103133 &  308.107754 &   44.042727 &  $-$1.173398 \\
      &       $\pm$.000886& $\pm$.00004124 &$\pm$.00013709 &$\pm$.000995 &$\pm$.000307 &$\pm$.000317 & $\pm$.000045 \\
&&&&&&& \\
n6  &  2019 Dec.\,8.737214&     2.01415693 &    3.36390123 &  209.097803 &  308.110505 &   44.041417 &  $-$1.173643 \\
      &       $\pm$.008107& $\pm$.00007506 &$\pm$.00035119 &$\pm$.002344 &$\pm$.000960 &$\pm$.000551 & $\pm$.000137 \\
&&&&&&& \\

JPL2 & 2019 Dec.\,8.758660&     2.01400668 &    3.36269842 &  209.103247 &  308.106996 &   44.043118 & $-$1.173133 \\
     &        $\pm$.000826& $\pm$.00002790 &$\pm$.00011365 &$\pm$.000816 &$\pm$.000223 &$\pm$.000255 &             \\
&&&&&&& \\
\hline 
\end{tabular}
}}
\end{table*}

\section{The comet orbit}\label{sect:comet_orbit}

All our attempts to obtain a barycentric original orbit of 2I/Borisow started with osculating orbit determination which includes selection and weighting of available observations. More details on our methods can be found for example in \cite{kroli_dyb:2017} and references therein. Next, each osculating heliocentric orbit was propagated numerically through our planetary system backwards up to the heliocentric distance of 250\,au, where it was transformed to the barycentric frame. In Table~\ref{table:comet_orbital_elements_original} we present original orbital solutions based on positional data that were selected and weighted according to Bessel criterion \citep[see][for more details]{krolikowska-sit-soltan:2009}.

Our first attempt to obtain this comet's orbit during this study was a slight update of the orbit presented in \cite{dyb-kro-wys-v1:2019}. It was performed on September~24, when 548 positional observations were available at the MPC database\footnote{https://minorplanetcenter.net/db\_search/} covering data-arc from 2019~Aug.~30 to 2019~Sept.~23. This preliminary orbital solution was named here (a6). From the osculating orbit, we obtained a barycentric original orbit for the first stage of our past motion studies. This starting original orbit is presented as the first row of Table~\ref{table:comet_orbital_elements_original}. 

More than a month later the data-arc and the number of positional observations increased considerably and we repeated the whole orbit determination process to observe how orbital elements were changed. We used 1459 observations spanning the interval: 2019~Aug.~30 -- 2019~Oct.~21 which were available at the MPC database on the 2019~October~30. We calculated a new orbit using this data, and elements and their uncertainties for this original barycentric orbit are given as solution (b5) in  Table~\ref{table:comet_orbital_elements_original}.  At that moment, non-gravitational (NG) effects were extremely uncertain and they could not improve the quality of the orbital fitting to this data-arc. For the sake of comparison, in Table~\ref{table:comet_orbital_elements_original} we also included  orbital solutions presented at that moment at MPC and Jet Propulsion Laboratory (JPL) Small-Body Database Browser\footnote{https://ssd.jpl.nasa.gov/sbdb.cgi}. In both cases we propagated their osculating orbits back in time to obtain a barycentric original orbit at a distance of 250\,au from the Sun. 

It is worth to note that the Bessel criterion of data selection \citep{krolikowska-sit-soltan:2009}, which we use here, rejected less than 5\% of observations, whereas  MPC and JPL solutions, retrieved on 2019~Oct.~31 and enclosed in Table~\ref{table:comet_orbital_elements_original}  were based on the similar data arc but used only about 85\% and less than 50\%  of available measurements, respectively. 

When comparing our first orbit (a6) with the solutions (b5) and (JPL1), one can easily note that a great increase in a number of positional data did not changed orbital parameters considerably, but reduced their uncertainty for about one order of magnitude . This is a very important fact for our investigation -- our knowledge of past motion of 2I/Borisov is quite good.  

Moreover, closer to the perihelion, the NG~activity of this object will probably increase, which might 'contaminate' orbit determinations from the point of view of the original orbit, see \cite{kroli-dyb:2012} for an extended discussion on data interval selection for original orbit determination.

We were aware that the only fact, that might considerably influence our knowledge on the past motion of 2I/Borisov, is the finding of prediscovery observations. And this happened on November~4, during writing of this paper. Nine prediscovery positions \citep{Bolin_et_al:2019} were made available at MPC, extending the observational arc backwards by over eight months. For this reason, we decided to repeat once more our orbit determination. 

As a result, we used 1711 observations available at MPC on October~5. They are distributed over the period of 324~days from 2018 Dec.~13 to 2019 Nov.~2 and in the range of heliocentric distance from 7.86\,au to 2.16\,au. There is a three month gap between 2019 May~5 and August~30, meaning no data between heliocentric distance from 5.09\,au to 2.99\,au.  Almost one year data arc allowed us to obtain the precise purely gravitational (GR) orbit, see solution (c5) in Table~\ref{table:comet_orbital_elements_original}. For the sake of comparison we also include an original orbit obtained from the solution presented at MPC on 2019 October~5 (MPC2).

We also attempted to determine NG~orbits, and two NG~solutions (with three NG~parameters, see Section~\ref{subsec: NG_orbits} for details) are also given in Table~\ref{table:comet_orbital_elements_original}: (n5) which is based on standard $g(r)$-function \citep{marsden-sek-ye:1973}, and (n6) -- based on $g(r)$-like function assuming CO~sublimation. All (c5), (n5), and (n6) solutions are based on weighted data. In Table~\ref{table:comet_orbital_elements_original} we also presented original NG~orbit, (JPL2), obtained from the osculating orbit available at JPL Small-Body Database Browser on November~5, where only one NG~parameter, $A_{\rm 2}$, was derived. This last NG~solution is based on analogous  $g(r)$-like function as our (c6) solution (see below). Uncertainties of orbital parameters of our both NG~orbits are about two times greater compared with GR~orbit (see Table~\ref{table:comet_orbital_elements_original}).

\subsection{Non-gravitational orbit of 2I/Borisov}
\label{subsec: NG_orbits}
We applied here a standard and widely used model of NG~forces proposed  by \cite{marsden-sek-ye:1973} to determine a NG-orbit from positional data. In this formalism, all three orbital components of the NG~acceleration acting on a comet are  proportional to the $g(r)$ function (symmetric relative to the perihelion):

\begin{eqnarray}
F_{i}=A_{\rm i} \> g(r),& A_{\rm i}={\rm ~const~~for}\quad{\rm i}=1,2,3,\nonumber\\
 & \quad g(r)=\alpha(r/r_{0})^{m}[1+(r/r_{0})^{n}]^{k},\label{eq:g_r}
\end{eqnarray}

\noindent where $F_{1},\, F_{2},$ and $F_{3}$ describe radial, transverse, and normal components of the NG~acceleration respectively. The radial component is defined as positive outward along the sun--comet line.

Table~\ref{table:comet_orbital_elements_original} shows parameters of two NG~orbits obtained by us using two different forms of the $g(r)$ function: the standard one describing the sublimation of water ice, solution (n5), and the function corresponding to the CO sublimation, solution (n6) (see Table~\ref{tab:gr-like_functions}). In both solutions we determined a set of three NG~parameters together with orbital elements. Therefore, our NG~solutions are characterized by slightly greater uncertainties of both orbital elements and NG~parameters $A_1, A_2$ and $A_3$ compared with the (JPL2) solution, where the existence of only one, transverse component of NG~acceleration, was assumed, see Tables~\ref{table:comet_orbital_elements_original} and \ref{tab:gr-like_functions}. It is worth to note that in case of neglecting normal component of NG~acceleration the negative $A_1$ parameter is obtained for both forms of $g(r)$-like function used by us, and also for the same $g(r)$-like function as in the case of (JPL2) solution. Taking into account two or three NG parameters, only slight decrease of RMS was obtained at the level of $10^{-2}$\,arcsec.

Moreover, in a standard model of NG~acceleration in a comet's motion, the parameter $A_1$ should be positive and should dominate all the others. Table~\ref{tab:gr-like_functions} shows that at least the last condition is not met, because in models (n5) and (n6) the radial and transverse component of NG~acceleration are comparable. In addition, in the model (n5) radial component is directed toward the Sun. Therefore, in our opinion, all our NG~solutions are very uncertain. However, we decided to check if they change the past close encounter parameters with the binary star Kruger\,60. 

\begin{center}
	\begin{table*}
		\caption{\label{tab:gr-like_functions}Parameters used in Eq.~\ref{eq:g_r} and NG~parameters for three NG~solutions presented in Table~\ref{table:comet_orbital_elements_original}.  NG~parameters $A_1$, $A_2$, and $A_3$ are given in units of $10^{-8}$\,au\,day$^{-2}$}
		\setlength{\tabcolsep}{6.0pt} 
		\begin{tabular}{ccccccccc}
			\hline 
            \\
            \multicolumn{9}{c}{standard g(r)-function (water ice sublimation) }\\
            \\
      Solution &   $\alpha$    & $r_0$ [AU] & $m$      & $n$   & $k$  & $A_1$  & $A_2$  & $A_3$         \\
	  n5       &	  0.1113      & 2.808 & $-$2.15  & 5.093 & $-$4.6142   & $-841.7 \pm 126.7$ &  $-672.1 \pm 113.0$ &  $-368.3 \pm 56.2$   \\
			\\
			\multicolumn{9}{c}{g(r)-like function (CO ice sublimation)} \\
			\\
	 Solution &  $\alpha$    & $r_0$ & $m$      & $n$   & $k$     & $A_1$  & $A_2$  & $A_3$       \\
	 n6         &     0.01003     & 10.0  & $-$2.0   & 3.0   & $-$2.6  & $11.389 \pm 5.095$ &  $-22.721 \pm 9.599$ &    $-8.398 \pm 2.765$    \\
	 JPL2       &     0.04084     & ~~5.0  & $-$2.0   & 3.0   & $-$2.6  & 0 (assumed) & ~~$25.838 \pm 4.616 $ & ~~0 (assumed)    \\
			\hline
		\end{tabular}
	\end{table*}
\end{center}

\section{Searching for home among nearby stars}
\label{sect:stars}

During our long-standing project devoted to the detailed study of the long period comets dynamics we collected a list of stars that are recognised as potential perturbers of cometary motion in the vicinity of the Sun. We started long ago with a few tens of stars \citep{dyb-kan:1999,dyb-hab3:2006}. The main limitation was a small number of available parallaxes and radial velocities.  Now the situation is much better due to the Gaia mission \citep{Gaia_mission:2016}. The Gaia DR2 catalogue \citep{Gaia-DR2:2018} provides us with the five-parameter astrometry for 1.3 billion stars, and a large subset of it has also radial velocities measured. We also use radial velocities from other sources.

At the moment we have finished updating our list of potential stellar perturbers basing mainly on the Gaia DR2 catalogue \citep{Gaia-DR2:2018}. This list contains 647 stars that visited in the past, are currently in or will visit in the future the solar vicinity and includes mass estimations found in the literature, more details can be found in \cite{RPM:2019}. This list will be made available to the public in the near future.

To search for a possible origin of 2I/Borisov we traced numerically its motion to the past as a N-body problem consisting of fully interacting: the comet, the Sun (in fact the solar system barycenter) and 647 stars or stellar systems. The starting data for a comet for such a calculation is a position and velocity in one of the original, barycentric orbits presented in Table~\ref{table:comet_orbital_elements_original}. For the numerical integration of equations of motion we used a framework described in detail in \citet[][DB15]{dyb-berski:2015}. In a Galactocentric model of motion all mutual interactions (i.e. star -- star attraction) were included as well as the action of the overall Galactic potential. Details of the calculation model, parameters, and methods can be found in DB15.

During the very first attempt at searching for the home system of 2I/Borisov we took an orbit obtained by Nakano and published by MPC on September 21 in CBET~4670, and numerically integrated the motion of 2I/Borisov, the Sun and 647 stars or stellar systems from our list of potential stellar perturbers of comet motion. 

As a result, in \citet[][v1]{dyb-kro-wys-v1:2019} we concluded that 1\,Myr ago 2I/Borisov passed double star Kruger\,60 at a distance of 1.74\,pc having an extremely small relative velocity of 3.43\,km\,s$^{-1}$. Repeating this later with an updated orbit of the comet gives almost the same result (Sect.~\ref{sect:past_proximity}).

\begin{figure}
\begin{center}
\includegraphics[width=0.46\textwidth]{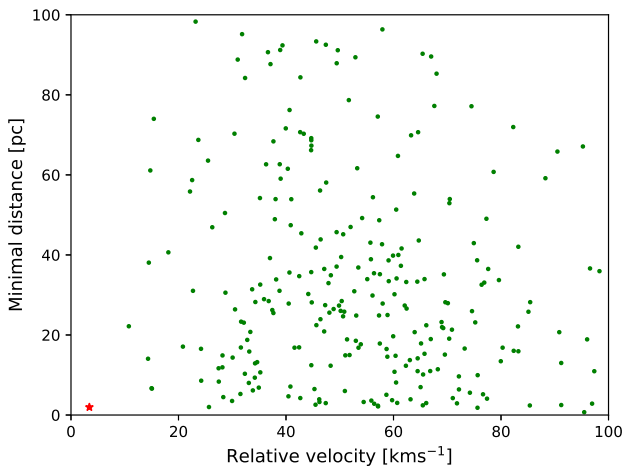}
\caption{Past encounters of 2I/Borisov with selected nearby stars. Cases with either miss-distance or relative velocity greater than one hundred are omitted here, leaving 264 data points. Kruger\,60 system is marked by a red star. The orbital solution (c5) is used here.}
\label{fig:stars}
\end{center}
\end{figure}

The overall picture of 2I/Borisov past encounters with stars from our list is presented in Fig.~\ref{fig:stars}. As many as 363 stars from our list have more or less close past approach with this comet. Our plot is limited to relative velocities below 100\,km\,s$^{-1}$ and minimal distance below 100\,pc so only 264 stars are included. In addition, parameters for all nine approaches below 2\,pc are presented in Table~\ref{tab:distances} in a chronological order; due to the limit in the relative velocity described above only four of them are included at the bottom of Fig.~\ref{fig:stars}.

The inspection of Table~\ref{tab:distances} shows, that during the last one million years, after a close approach with Kruger\,60, 2I/Borisov made eight close passages near low mass stars at a high relative velocity before it entered the Solar System. Taking into account the minimal distance and the estimated mass GJ~234~AB system seems to be the most important.  Our knowledge on the GJ~234~AB system's kinematics is rather poor but a moderately high relative velocity rather excludes this system from a list of potential home systems of 2I/Borisov. Nevertheless, some small deflection of the 2I/Borisov path by GJ~234~AB is possible. This is taken into account in our calculations since we use a N-body model with all mutual interactions included. However, the uncertainty of the parameters of this system is a source of an additional uncertainty of 2I/Borisov past trajectory. How it increases the uncertainty of parameters of the comet's meeting with the Kruger\,60 is difficult to estimate without very extensive numerical calculations. A better knowledge on GJ~234~AB system kinematics and masses would be desirable to refine Kruger~60 approach parameters.

Basing on the results presented in Table~\ref{tab:distances}  we found, that our conclusion presented in \citet[][v1]{dyb-kro-wys-v1:2019} was correct and Kruger~60 is the only acceptable candidate for a home system of 2I/Borisov from among all 647 stars in our list.

	\begin{table*}
		\caption{\label{tab:distances} Parameters of all close encounters below 2\,pc with stars in our list. The rows are presented in a reversed chronological order. In consecutive columns we present: a star name, epoch of the encounter (t), a miss-distance (ds), relative velocity (vrel), a heliocentric distance of the encounter (dh) and an estimated mass of the star.}
		\begin{center}
		\begin{tabular}{lcccccccccc}
		\hline
		Star name & t [Myr] & ds [pc] & vrel [km\,s$^{-1}$] & dh [pc] & mass [M$_{\odot}$ \\
		\hline
   Kapteyn's star            &  -0.013 &    1.965 &  269.748 &  0.415 & 0.30 \\
   van Maanen's Star         &  -0.015 &    0.648 &  284.117 &  0.492 & 0.50 \\
   GJ 4274                   &  -0.023 &    0.587 &  309.357 &  0.763 & 0.14 \\
   GJ 1111                   &  -0.040 &    1.816 &   75.611 &  1.325 & 0.12 \\
   WISEA J104335.09+121312.0 &  -0.053 &    0.932 &  269.534 &  1.744 & 0.08 \\
   2MASS J05565722+1144333   &  -0.079 &    1.765 &  175.893 &  2.594 & 0.12 \\
   GJ 1103 AB                &  -0.095 &    0.707 &   95.476 &  3.131 & 0.18 \\
   GJ 234 AB  (Ross 614)     &  -0.137 &    1.998 &   25.656 &  4.527 & 0.34 \\
   Kruger 60                 &  -0.998 &    1.970 &    3.418 & 32.962 & 0.44 \\

			\hline
		\end{tabular}
		\end{center}
	\end{table*}

\begin{figure*}

\includegraphics[width=0.49\textwidth]{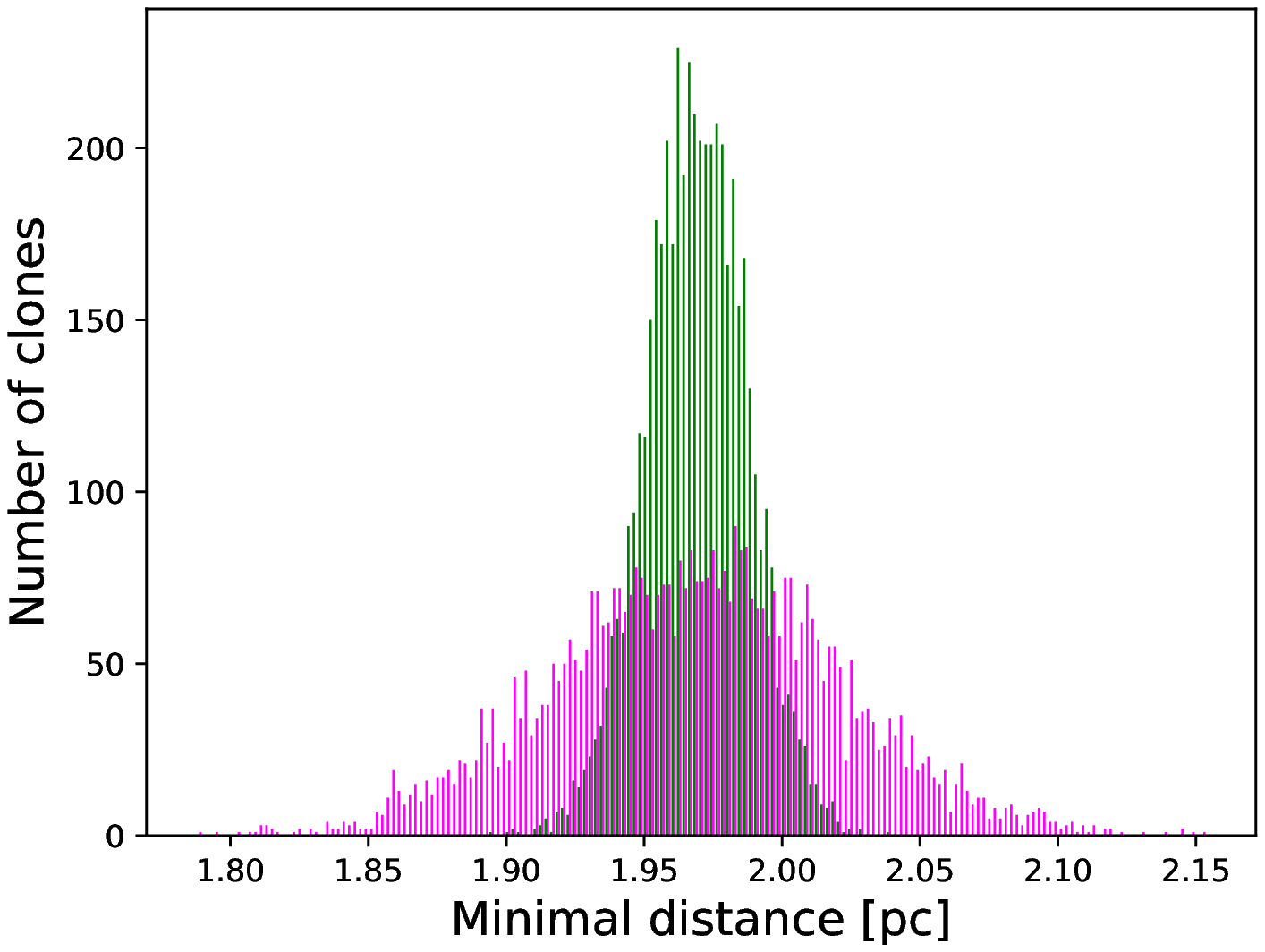}
\includegraphics[width=0.49\textwidth]{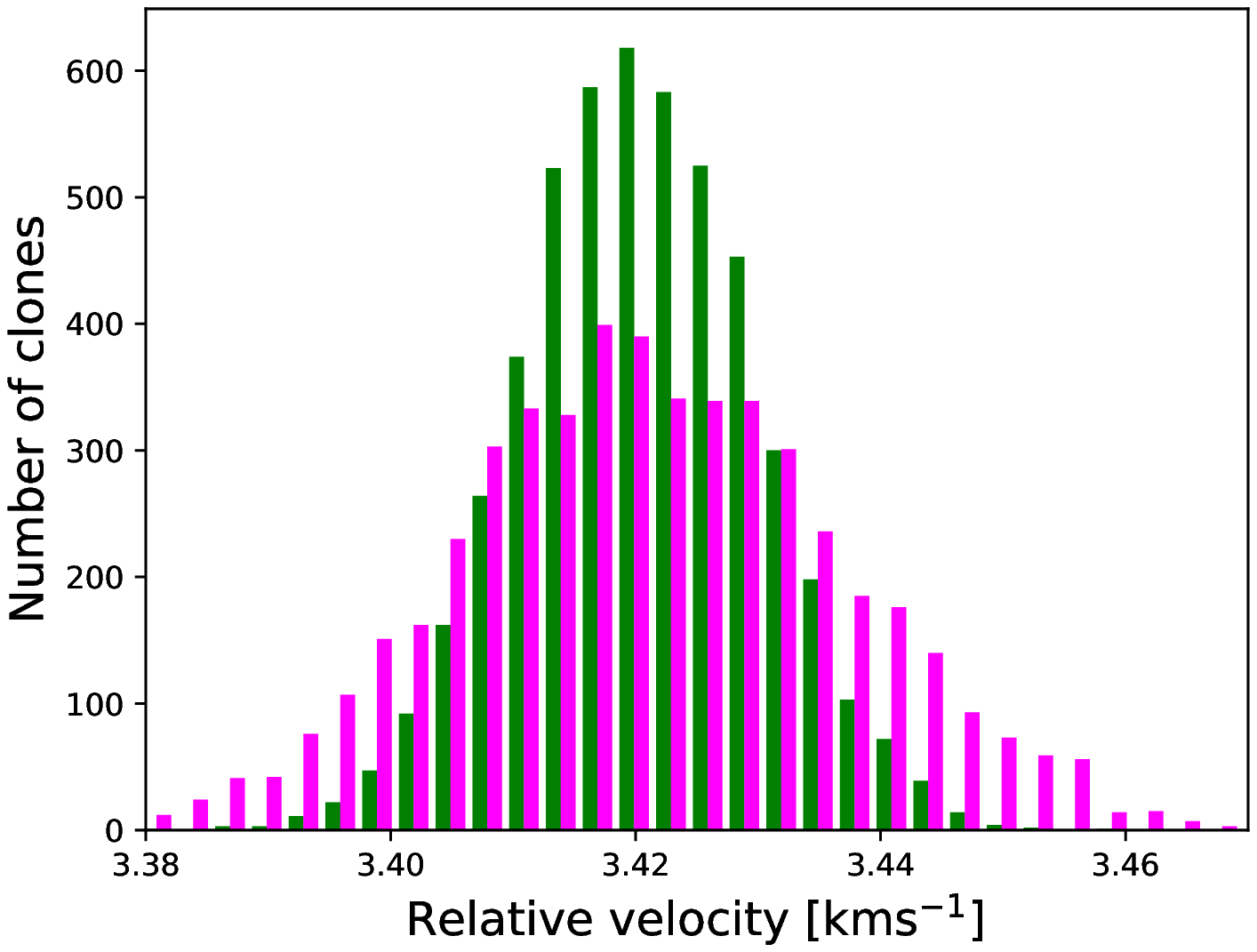}

\includegraphics[width=0.49\textwidth]{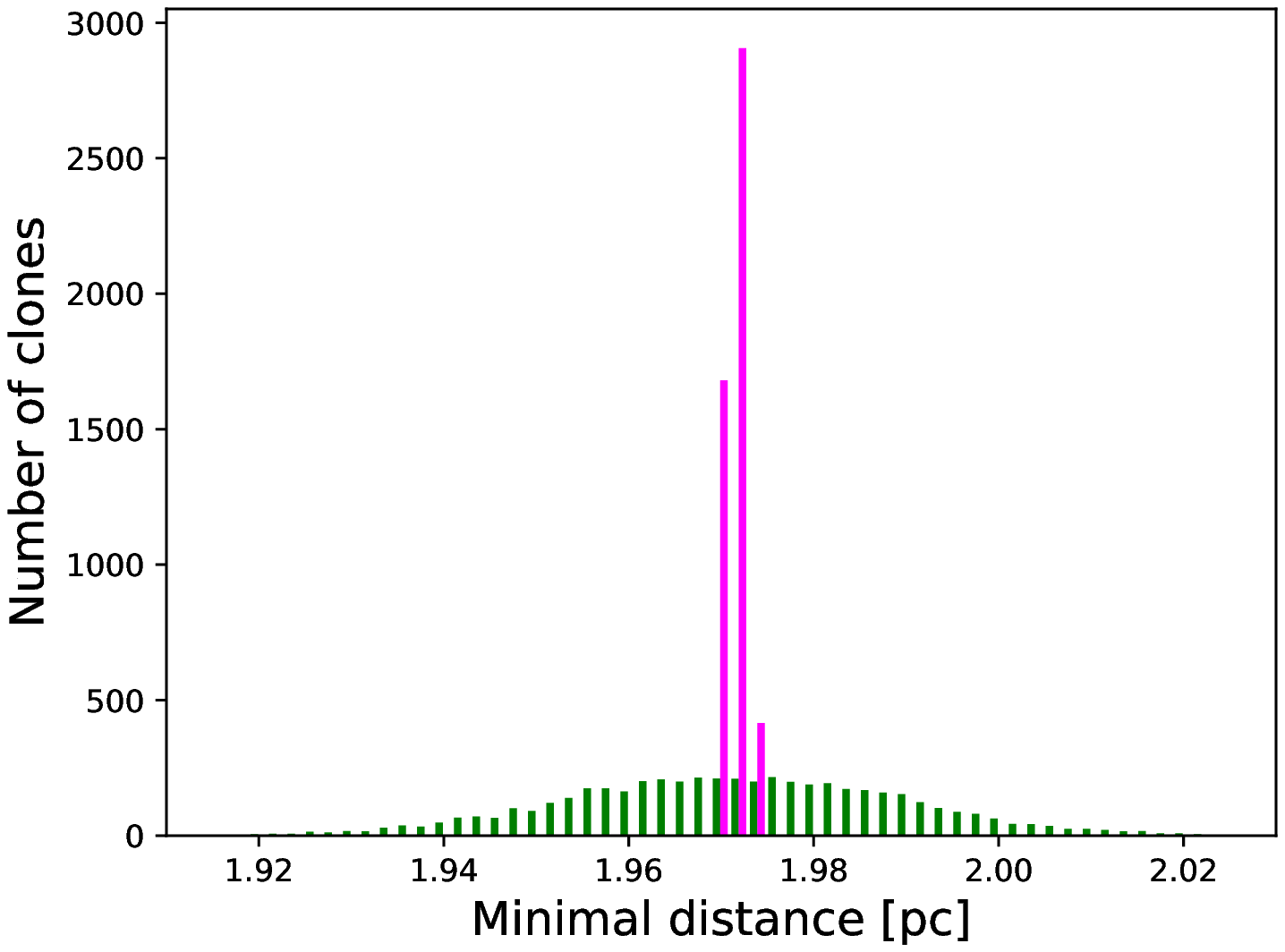} 
\includegraphics[width=0.49\textwidth]{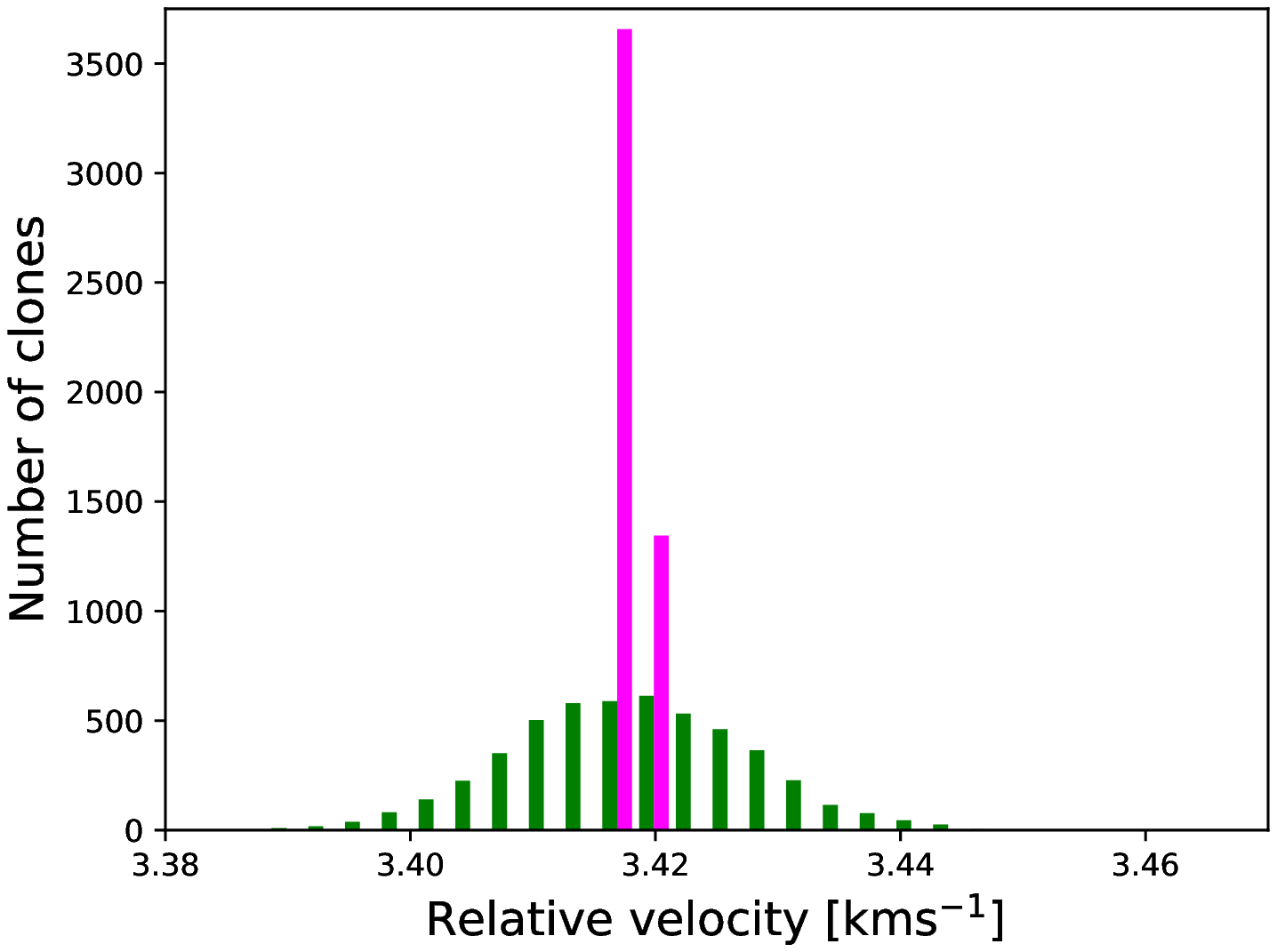} 

\includegraphics[width=0.49\textwidth]{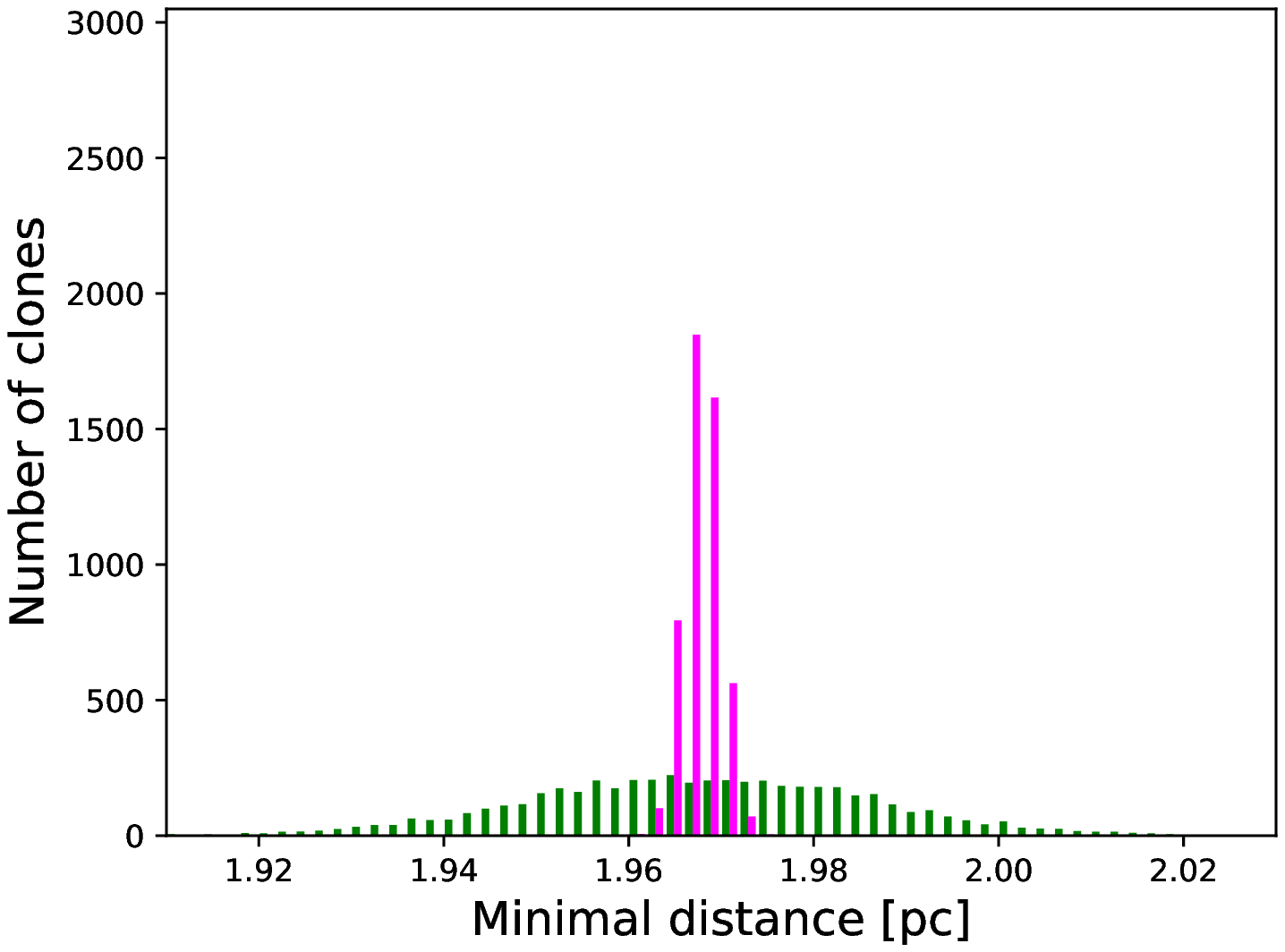}
\includegraphics[width=0.49\textwidth]{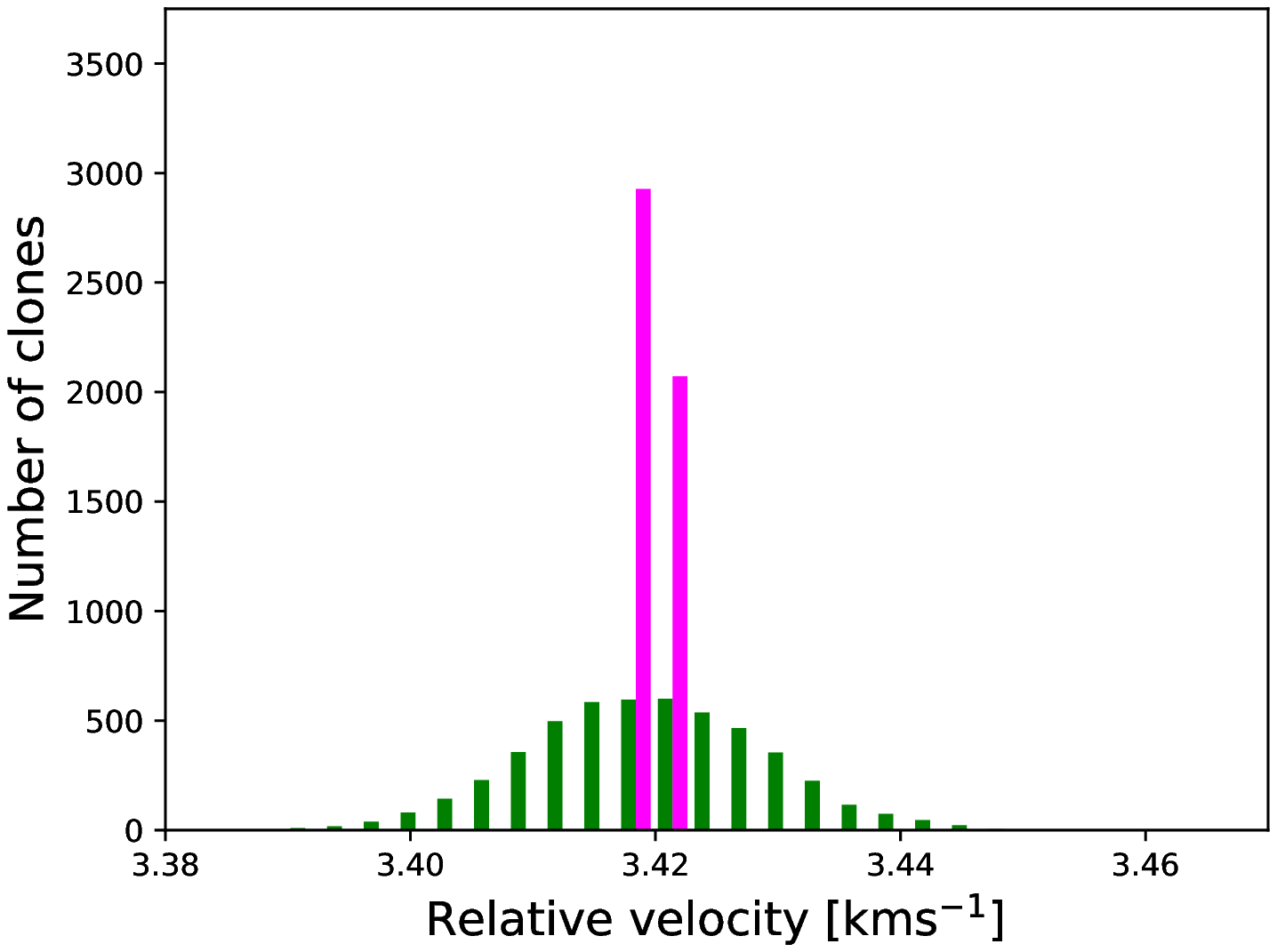}
   
    \caption{ Left column: the distribution of minimal distances between 2I/Borisov and the Kruger\,60 system. Right column: the distribution of relative velocities at the time of the closest encounter between 2I/Borisov and the Kruger\,60 system. The bin width are the same across the whole column. The upper row is for (a6) solution, the middle row for (c5) and the bottom one for (n6). In all histograms green bars indicate close encounters of the clones of the star with the comet in its nominal orbit, whereas magenta bars designate values obtained from examining the clones of the comet motion and the nominal star. }
    \label{fig:histogramy_new}
\end{figure*}

\section{The binary star Kruger 60}
\label{sect:Kruger}

A well known visual binary Kruger~60 is named after Adalbert Krüger who observed it in 1873. Both components are of M spectral type and circle each other in an orbit with a period of 44.6\,years and an eccentricity of 0.41. The components passed their periastron about five years ago according to \cite{Soderhjelm:1999}. Other designations include BD+56~2783, GJ\,860AB, HD\,239960, HIP\,110893, and ADS\,15972. This is a tenth closest multiple stellar system, currently only 4\,pc from the Sun and is approaching to us. The multiplicity of the system is extensively analysed in \cite{Knapp:2018}.   It is also worth to mention that this star is the subject for exoplanet search, see for example \cite{Helminiak:2009, Bonavita:2016, Butler:2017}.

Recently both components of Kruger~60 were observed by Gaia mission spacecraft and the most precise astrometry at the moment is available in the Gaia DR2 catalogue \cite{Gaia-DR2:2018}, objects: Gaia\,DR2\,2007876324466455424 and Gaia\,DR2\,2007876324472098432, but radial velocities are missing in this source.

\begin{table*}
		\caption{\label{tab:Kruger_parameters} Parameters of Kruger~60 system used in this paper.}
		\begin{center}
		\begin{tabular}{lccr}
		\hline
		&&& \\
		Parametr & Value & Uncertainty & Source \\
		&&& \\
		\hline
		&&& \\
		\multicolumn{4}{c}{Component A} \\
		&&& \\
		Barycentric right ascension (ICRS) at Ep=2015.5 & 336\fdg99231372434 & 0.1817 [mas] & Gaia DR2 \\
		Barycentric declination (ICRS) at Ep=2015.5 & +57\fdg69406039042 & 0.1329 [mas] & Gaia DR2 \\
		Absolute parallax [mas] & 249.3926 & 0.1653 & Gaia DR2 \\
		Proper motion in right ascension [mas/yr] & $-725.227$ & 0.537 & Gaia DR2 \\
		Proper motion in declination [mas/yr] & $-223.461$ & 0.348 & Gaia DR2 \\
		Mass [M$_{\odot}$] & 0.270  & estimated & \cite{Bonavita:2016} \\
		Radial velocity [km\,s$^{-1}$] & $-24.0$ & 5  & \citep{wilson:1953} \\
		&&& \\
		\multicolumn{4}{c}{Component B} \\
		&&& \\
		Barycentric right ascension (ICRS) at Ep=2015.5 & 336\fdg99162241967 & 0.6797 [mas] & Gaia DR2 \\
		Barycentric declination (ICRS) at Ep=2015.5 & +57\fdg69419743802 & 0.6711 [mas] & Gaia DR2 \\
		Absolute parallax [mas] & 249.9668  & 0.7414 & Gaia DR2 \\
		Proper motion in right ascension [mas/yr] & $-934.098$ & 1.319 & Gaia DR2 \\
		Proper motion in declination [mas/yr] & $-686.244$ & 1.410 & Gaia DR2 \\
		Mass [M$_{\odot}$] & 0.180 & estimated & \cite{Bonavita:2016} \\
		Radial velocity [km\,s$^{-1}$] & $-28.0$  & 5  & \citep{wilson:1953} \\
		&&& \\
		\hline
		\end{tabular}
		\end{center}
\end{table*}

Individual radial velocities of Kruger~60 components can be found in the SIMBAD database\footnote{http://simbad.u-strasbg.fr/simbad}, which quotes values from the General Catalogue of Stellar Radial Velocities \citep{wilson:1953}, then repeated in \cite{WEB_cat:1995}. In the original catalogue they estimate the mean errors of these values as 2.5\,km\,s$^{-1}$ maximal error as 5\,km\,s$^{-1}$ and this last value is quoted in SIMBAD. These radial velocities are: $-24.0$\,km\,s$^{-1}$ for HD\,239960A and $-28.0$\,km\,s$^{-1}$ for HD\,239960B. Both were measured at the Mount Wilson observatory in the first half of the 20$^{th}$ century but the epochs of observations are not given.

It is worth to mention, that there are several other radial velocity measurements for these stars in the literature, ranging from $-16$ to $-36$~km\,s$^{-1}$ but in many cases without clear statement which component was observed or a value for only one component is offered. For example, in \cite{Sperauskas:2016}, one can found the value of $-31.5$\,km\,s$^{-1}$ but for the object HIP\,110893, which is the name of the Kruger~60 system as a whole. 

There is an additional problem resulting from the lack of radial velocities in the Gaia DR2 catalogue. Kruger~60 has a relatively short orbital period of 44.6\,years therefore to reconstruct spatial positions and velocities of its components we should use positions, proper motions and radial velocities determined for the same epoch. At the moment it is impossible.

For the investigation presented in Section~\ref{sect:past_proximity} we calculated the center of mass position and velocity using Gaia DR2 positions, proper motions and parallaxes augmented with individual radial velocities found in \cite{WEB_cat:1995} and using mass estimations of 0.270 M$_{\odot}$ and 0.180 M$_{\odot}$ from \cite{Bonavita:2016}, all parameters adopted for the Kruger~60 system are listed in Table~\ref{tab:Kruger_parameters}.

\section{The past proximity} \label{sect:past_proximity}

To confirm the possibility that Kruger~60 might be a home system for 2I/Borisov and assess the uncertainty of our finding, we decided to determine our own orbital solutions for 2I/Borisov (see Section \ref{sect:comet_orbit}) which allowed us to generate 5000~clones of this object according to the method proposed by \cite{sitarski:1998}. Additionally, to estimate the influence of the uncertainty of stellar parameters on our result, we also generated 5000~clones of the double star Kruger\,60 with the help of a covariance matrix enclosed in the Gaia DR2 catalogue. Due to the lack of radial velocity uncertainties and some discrepancies in the literature, we decided not to vary this parameter at the first stage of this study. Next, we integrated all comet clones with the nominal star and all star clones with the nominal comet.

The results for orbital solutions (a6), (c5) and (n6) are shown in Fig.\ref{fig:histogramy_new}. These histograms visualize the influence of stellar data inaccuracy and comet orbit uncertainties on the encounter parameters. We present distribution of minimal distances between considered bodies in the left column and their relative velocity distributions in the right one. In the first row, where (a6) orbit is used, it can be observed that green histograms which present results of the integration of the star clones with the nominal comet orbit are much more compact than the magenta histograms showing results of the integration of the clones of the comet and the nominal stellar data (note a different horizontal scale in the left upper plot). {\bf This led us to the conclusion that, in this case, with data available at that moment, our knowledge on the space motion of the comet is more uncertain than on the motion of the star}.

After the prediscovery observations of 2I/Borisov were announced we repeated  the whole orbit determination procedure, as it is described in Section~\ref{sect:comet_orbit}. We conclude that our pure GR~solution (c5) is the best one for 2I/Borisov past motion study because in our opinion the NG~orbit determinations of this comet are very uncertain at the moment, see Section~\ref{subsec: NG_orbits} for more details. 

The histograms for the (c5) solutions are presented as a middle row of Fig.~\ref{fig:histogramy_new}. The nominal values of the minimal distance and the relative velocity during close encounter between the comet and Kruger~60 remain almost the same (see Table~\ref{tab:Kruger_approach}) but a considerable change in the histograms appeared (note the change in a horizontal scale of the minimal distance histogram): the distributions for the comet clones (in magenta) became very compact in comparison with the green histograms for the star clones. It means that now the uncertainty of close approach parameters are highly dominated by the stellar data error estimates.

For the completeness, in the lower row of Fig.~\ref{fig:histogramy_new} we also present distributions based on the NG-solution (n6). Table~\ref{tab:Kruger_approach} shows that two our NG~solutions presented in Table~\ref{table:comet_orbital_elements_original} give similar results compared with (c5) orbit, except the uncertainties which are about two times larger resulting in slightly more dispersed magenta histograms in the lower row of Fig.~\ref{fig:histogramy_new}.

However, in our opinion, these slightly greater uncertainties might better reflect our knowledge of the past journey of 2I/Borisov among the stars.  

{\bf We conclude here, that while our main result announced in \cite{dyb-kro-wys-v1:2019} remains almost unchanged: a minimal comet -- star distance equals 1.97\,pc at the relative velocity equals 3.42\,km\,s$^{-1}$ (see Table~\ref{tab:Kruger_approach}), the uncertainty source arrangement reversed: now we know the motion of 2I/Borisov with a much higher accuracy than the path of the star. } 

For the sake of comparison in Table~\ref{tab:Kruger_approach} we included nominal parameters of the close approach of 2I/Borisov with Kruger~60 system for all considered orbital solutions. We also added these parameters for orbits published at MPC and JPL see Table~\ref{table:comet_orbital_elements_original}. One can see that all parameters of close encounter between nominal orbit of 2I/Borisov and nominal orbit of Kruger\,60 are strikingly similar.

\begin{table}
		\caption{\label{tab:Kruger_approach} Parameters of close encounters of nominal orbit of 2I/Borisov with nominal space trajectory of Kruger~60 center of mass for different orbital solutions included in Table~\ref{table:comet_orbital_elements_original}. In consecutive columns we present the orbital solution code, epoch of the encounter (t), a miss-distance (ds), relative velocity (vrel) and a heliocentric distance of the encounter (dh).}
		\begin{center}
		\begin{tabular}{lcccc}
		\hline
		Orbit & t [Myr] & ds [pc] & vrel [km\,s$^{-1}$] & dh [pc] \\
		\hline
a6    & -1.019 &  1.970 &  3.420 &  33.63\\
b5    & -0.989 &  1.954 &  3.424 &  32.66\\
MPC1  & -0.966 &  1.990 &  3.413 &  31.89\\
JPL1  & -0.980 &  1.974 &  3.417 &  32.36\\
c5    & -0.974 &  1.972 &  3.418 &  32.16\\
MPC2  & -1.022 &  1.986 &  3.415 &  33.72\\
n5    & -1.022 &  1.971 &  3.418 &  33.76\\
n6    & -1.017 &  1.968 &  3.420 &  33.59\\
JPL2  & -0.993 &  1.988 &  3.412 &  32.78\\
			\hline
		\end{tabular}
		\end{center}
\end{table}

\subsection{Radial velocity uncertainty of Kruger~60}

{\bf It is obvious, that a minimal distance of  1.97\,pc cannot be treated as an evidence for Kruger~60 being a source of 2I/Borisov -- it is far to big}. But there is one large source of the uncertainty in our result not studied so far -- this is the lack of precise radial velocities for Kruger~60 components. In all our calculations described above we used radial velocities obtained from the SIMBAD database, which are old and inaccurate determinations. Combining these velocities with positions, proper motions and a parallax from Gaia DR2 gives only a rough information on this system kinematics.

To explore the influence of radial velocity uncertainties of Kruger~60 components  on the parameters of the close approach of this system with 2I/Borisov we repeated a generation of 5000~clones of the Kruger\,60 system, now drawing only the radial velocity of the center of mass of Kruger~60 from an uniform distribution spanning the interval from 15 to 35\,km\,s$^{-1}$. 

\begin{figure}
\begin{center}
\includegraphics[width=0.48\textwidth]{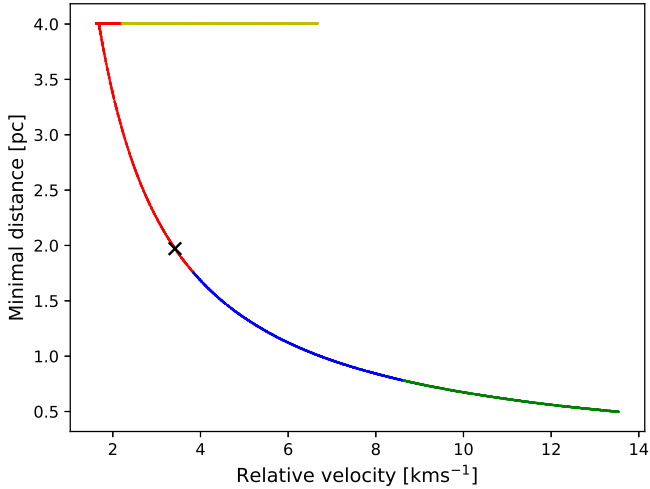}
\caption{Past encounter parameters of 2I/Borisov (using (c5) solution) with Kruger\,60 system clones derived from the Gaia DR2 astrometry with random systemic radial velocity. The radial velocity of Kruger~60 is color-coded (see text) and the black cross marks the nominal value.}
\label{fig:clones}
\end{center}
\end{figure}

The distribution of the close approach parameters resulting from this simulation is presented in Fig.~\ref{fig:clones}. The colours of dots correspond to the systemic radial velocity used (all in km\,s$^{-1}$): yellow for $-35<v_r<30$, red for $-30<v_r<25$, blue for $-25<v_r<20$ and green for $-20<v_r<15$. The horizontal line in the upper part of the plot (yellow and partly red) correspond to the geometry where there is no past approach between 2I/Borisov and Kruger~60 -- the distance of 4\,pc equals their current separation. One can note, that only points from the green interval are consistent with the hypothesis that 2I/Borisov came from Kruger~60. These points however correspond to higher relative velocities. 

We know that long period comets are continuously ejected by planets from the Solar System into an interstellar space with heliocentric velocities of 3 -- 5\,km\,s$^{-1}$. But in case of 2I/Borisov larger escape velocity is physically quite possible remembering that Kruger\,60 is a double star. 

\section{Conclusion and prospects}
\label{sect:conclusions}

We have inspected a list of almost 650 nearby stars or star systems as candidates for the home of the interstellar comet 2I/Borisov.  A paper describing our list of stellar perturbers of long period comet motion is under preparation an the list will be made publicly available in a near future.

In the first version of this note we concluded that the double system Kruger~60 is a possible source of this comet. While the obtained relative velocity at their past encounter was very low, their minimal distance was a bit too big. But we showed that it might become much smaller if we use much more accurate radial velocities of Kruger~60 components, unavailable at that moment.

Later on, with the updated and refined 2I/Borisov orbit we confirmed that result but the constraints on Kruger~60 radial velocity, not taken into account in the first version of this note, now made this candidate considerable less probable. 

Our current knowledge on 2I/Borisov original orbit seems to be quite good and new positional data for this comet taken around perihelion and at outgoing leg of orbit will not allow for more accurate examination of the past approaches of this comet to Kruger~60 and other stars. However, they will probably allow to better determine the NG~orbit of this comet, and eventually examine its future journey beyond the Solar System. It should be stressed that for now the uncertainty of the past close approach parameters of 2I/Borisov with Kruger~60 are highly dominated by the (small !) errors of this double star astrometry even in the case of assuming zero error in radial velocities. 

The small, additional but difficult to estimate source of the uncertainty of our result, is the lack of the knowledge on kinematics and masses of the GJ~234~AB  (Ross 614) system which was passed some 140 kyrs ago by 2I/Borisov on its way to us.

During the writing of this updated note we have obtained a new, unpublished systemic radial velocity of Kruger~60 system equal to -34.07\,km\,s$^{-1}$ with the estimated error on the level of 0.5\,km\,s$^{-1}$ (Fabo Feng, private communication). It is obtained using the PEXO package \citep{FaboFeng:2019} basing on data from HIPPARCOS \citep{hip2_book:2007}, WDS catalogue \citep{WDS:2001} and LCES HIRES/Keck survey \citep{Butler:2017}. This systemic radial velocity corresponds to the yellow interval in Fig.~\ref{fig:clones} so the hypothesis that 2I/Borisov came from Kruger~60 seems to be completely ruled out. 

\section*{Acknowledgement}
We would like to thank Fabo Feng for discussing some aspects of this research and providing the new, unpublished radial velocity for Kruger~60. This research was partially supported by the project 2015/17/B/ST9/01790 founded by the National Science Centre in Poland.

\bibliographystyle{mnras.bst}

\bibliography{moja26}

\label{lastpage}

\end{document}